\documentstyle[preprint,revtex]{aps}

\def\phys{Phys. Rev. B}
\def\lett{Phys. Rev. Lett.}

\def\NH{\rm N_H}

\begin{document}
\draft

\begin{title}
Optical conductivity of the Hubbard model \\
at finite temperature
\end{title}

\author{Jos\'e ~A.~Riera$^1$ and Elbio ~Dagotto$^2$}

\begin{instit}
$^1$Center for Computationally Intensive Physics,   \\
Physics Division, Oak Ridge National Laboratory,
Oak Ridge, TN 37831         \\
and Department of Physics and Astronomy,      \\
Vanderbilt University,
Nashville, TN 37235.      \\

$^2$Department of Physics,
National High Magnetic Field Laboratory,      \\
Florida State University,
Tallahassee, FL 32306.
\end{instit}


\begin{abstract}
The optical conductivity, $\sigma(\omega)$, of the two dimensional
one-band Hubbard
model is calculated at finite temperature using exact
diagonalization techniques on finite clusters. The in-plane d.c.
resistivity, $\rho_{ab}$, is also evaluated. We find that
at large U/t and temperature T, $\rho_{ab}$ is approximately
linear with temperature, in reasonable agreement with
experiments on high-T$_c$ superconductors. Moreover,
we note that $\sigma(\omega)$ displays charge excitations, a
mid-infrared (MIR) band and a Drude peak, also as observed
experimentally.
The combination of the Drude peak and the MIR oscillator strengths
leads to a conductivity that decays slower than $1/\omega^2$ at
energies smaller than the insulator gap near half-filling.
\end{abstract}



\newpage

Experimentally, it has been observed that the in-plane
d.c. resistivity, $\rho_{ab}$, of the hole-doped high temperature
superconductors is linear with
temperature when the hole doping fraction is optimal, i.e. when
the critical temperature (T$_c$) is maximum.\cite{batlogg}
This simple phenomenological law is still one of the most
puzzling features of the normal state of the cuprates. A
possible explanation of this behavior using the Bloch-Gr\"uneisen
formula (which is based on electron-phonon scattering)
seems unlikely,\cite{ito} and thus mechanisms based on
scattering by spin fluctuations have been proposed.
The a.c. conductivity, $\sigma(\omega)$, also presents
interesting features. A mid-infrared band
(MIR) has been observed inside the charge-transfer gap of the
insulating parent compound.\cite{uchida} In addition,
at small frequency (relative to the gap) $\sigma(\omega)$
decays as $1/\omega$, instead of the more standard Drude
behavior $1/\omega^2$. This effect can be phenomenologically
described by an energy-dependent lifetime
$\tau(\omega) \sim \omega^{-1}$.
Several theoretical mechanisms have been proposed to explain
these features.\cite{review}
Most are based on simple mean-field solutions of electronic
Hubbard or t-J like models, but the validity of these
approximate descriptions is unclear. An alternative approach
involves direct numerical analysis of these models.
Recently, there has been considerable progress
in this approach and several studies of
$\sigma(\omega)$ on finite clusters using exact diagonalization
techniques (at $zero$ temperature) have been reported and
compared to analytical approaches.\cite{prelovsek,dago92}
The presence of the MIR band has been explained as due to
the considerable spectral weight located in the incoherent
part of the hole spectral function, and the anomalous
$1/\omega$ decay was attributed to a combination of oscillator
strength between the MIR band and the zero frequency Drude
peak at
zero frequency in the metallic regime.\cite{review}

In this paper we report a numerical
study of the two dimensional (2D) one-band Hubbard model
at $finite$ temperature using
the exact diagonalization approach on small clusters.
Little work has been carried out previously at nonzero
temperature using this technique, since the full set of
eigenvalues and eigenvectors of the finite cluster
is needed to determine thermal properties.
This substantially increases the memory and CPU
requirements relative to zero temperature properties.\cite{mc}
Here we evaluate both $\sigma(\omega)$ and $\rho_{ab}$ and
compare the results with experiments; we find encouraging
qualitative agreement.
The calculation of transport properties of a weakly dissipative
system in the context of many-body problems generally follows
the Kubo formulation, which relates the conductivity
to a current-current correlation function. This approach has
been widely used in the context of Hubbard-like models to
describe strongly correlated systems. The real part
of the conductivity at finite temperature is given by
\begin{eqnarray}
\sigma(\omega) = \pi {{(1 - e^{-\beta \omega} )}\over{\omega Z}}
\sum_{n,m} e^{- \beta E_n } | \langle n | j_x | m \rangle |^2
\delta ( \omega + E_n - E_m ),
\end{eqnarray}
\noindent
where $| n \rangle $ is an eigenstate of the Hubbard
Hamiltonian with eigenvalue $E_n$, $Z$ is the partition function,
$\beta$ the inverse of the temperature, and $j_x$ the current
operator in the x-direction. The rest of the notation is
standard, and details can be found in textbooks.\cite{mahan}

The diagonalization of the Hubbard model was carried out on
small square clusters. In each subspace corresponding to a
given set of quantum numbers (momentum, z-component of the
total spin and parity under spin reversal) we computed all
the eigenvalues and eigenvectors in two steps. First,
the matrix was reduced to a tridiagonal form using the Householder
algorithm. We then diagonalized the resulting matrix using a
standard QL algorithm. Since in general we have to deal with
complex hermitian matrices,
we developed hermitian versions of the subroutines TRED2
and TQLI of the Numerical Recipes package.\cite{press}
In principle, the total operation count for both subroutines
scales as $\sim \NH^3$, where $\NH$ is the dimension of the
matrix to be diagonalized. However, since the innermost loops
could be vectorized, we found that the coefficient of the
$\NH^3$ was four orders of magnitude smaller than the
coefficient of $\NH^2$ term on a Cray YMP.
The total CPU time required to diagonalize
the largest matrix with $\NH=540$ was approximately
23 seconds on a Cray YMP supercomputer and the total
memory required for the diagonalization of a ${\NH \times \NH}$
matrix was $ {\rm 2 \times \NH^2 + 4 \times \NH }$ words.
Both CPU-time and memory requirements compare well with similar
subroutines in other packages such as IMSL or NAG.
The calculation of $\sigma(\omega)$ itself was considerably more
CPU time consuming than the diagonalization procedure.
According to Eq. (1), the total operation count in a subspace
of dimension $\NH$ and for a fixed temperature
scales as $\NH^4$. However, in the calculation of the matrix
${\rm < n | j_x | m > } $ one can take advantage of its
sparse nature, thus effectively reducing the dependence
from $\NH^4$ to $\NH^3$.
Moreover, a set of measurements at different temperatures
could be done with almost the same CPU time as
a single temperature by appropriately rearranging the loops
and vectorizing the innermost one.

Although the technique applied here works equally well for the
Hubbard and t-J models, we have concentrated only on the Hubbard
model, which possesses excitations across the gap that are
important for comparison of $\sigma(\omega)$ with experiments.
These calculations were carried out on small square clusters
of eight and ten sites, similar to those
used previously in the study of the Heisenberg model and other
systems.\cite{oitmaa,review} On finite systems it is important
to choose the boundary conditions to appropriately minimize
finite size effects.
In the present study we decided to use antiperiodic boundary
conditions (APBC).\cite{twisted}
In the 8-site cluster with APBC the non-interacting limit
${\rm U/t=0}$ has 4 levels with energy -2t, and another
4 levels with energy +2t. In the half-filled case, and with
${\rm U/t > 0}$, two bands exist separated by a gap which
grows as ${\rm U/t}$ increases. This behavior is also expected
for the Hubbard model in the bulk limit.
When holes are introduced, they are energetically favored to
appear in the lower band.
In contrast, the same cluster with periodic boundary
conditions (PBC) contains 6 levels of zero energy at
${\rm U/t=0}$, one state with energy -2t and another with energy
+2t. The large zero-energy degeneracy appears to produce large
finite-size artifacts in the PBC case at finite coupling,
and for this reason APBC will be used in this paper.

The application of Eq.(1) to Hubbard-like models involves some
complications. One problem is that an isolated, finite system,
such as the clusters analyzed in any computational study,
cannot show resistive behavior. Thus, the resistivity of a
metallic ground state at zero temperature must vanish,
since a Drude-like weight $\rm D \delta(\omega)$ appears
at zero frequency in the conductivity. In fact, Eq.(1)
implies that a $\delta$-function is always present at zero
frequency if any eigenstate of the Hamiltonian satisfies
$| \langle n | j_x | m \rangle | \neq 0$ with $E_n = E_m$.
Therefore,
electron-electron interactions (umklapp processes) are not
sufficient to produce dissipation, and the d.c. resistivity
of the Hubbard model is zero at
all temperatures.\cite{giamarchi}
On finite systems with twisted BC the calculation of
the Drude weight is carried out
indirectly, using the two dimensional sum rule
\begin{eqnarray}
\int^{\infty}_0 d\omega \sigma(\omega) = {{\pi e^2}\over{4N}}
\langle - {\hat T} \rangle,
\end{eqnarray}
\noindent
where $\langle {\hat T} \rangle$ is the thermal average of the
kinetic energy operator. Assuming the existence of a
contribution ${\rm D} \delta(\omega)$ at zero energy, we obtain
\begin{eqnarray}
{{D}\over{2 \pi e^2}} = {{\langle - {\hat T} \rangle} \over{4N}}
- {{1}\over{\pi e^2}} \int^{\infty}_{0^+} d\omega
\sigma(\omega),
\end{eqnarray}
\noindent
where both terms on the r.h.s. of Eq.(3) can be calculated
numerically. Following this procedure, the inverse of the
Drude weight $\rm D^{-1}$ is plotted in Fig.1 as a function of
temperature for several couplings ${\rm U/t}$, and at a filling
of 6 electrons on the 8 site cluster
($\langle n \rangle = 0.75$). $\rm D^{-1}$ is
proportional to the d.c. resistivity, once a finite width is
given to $\delta(\omega)$ to mimic dissipative processes not
included in the Hamiltonian. It is interesting to
note that for ${\rm T > t}$ and strong coupling, $\rm D^{-1}$
is approximately linear with temperature. This is in agreement
with the predictions of Rice and Zhang\cite{rice} for the
large U/t limit. On reducing the Hubbard coupling ${\rm U/t}$,
we find that $\rm D^{-1}$ acquires curvature in T, and in
the weak coupling region $\rm D^{-1} \sim T^{2}$. This
behavior is consistent with the quadratic temperature
dependence of the resistivity expected for a Fermi liquid.
Now we will consider how our results can be compared
with experiment,\cite{batlogg}. First,
note that for ${\rm t = 0.4 eV}$ a temperature of $1t$
corresponds to
approximately $4600 \rm K$ which is much higher than the
maximum experimental temperature for $\rho_{ab}$ of
$\approx 800 \rm K$. In principle we should reduce the
temperature in our cluster calculations for this
comparison. Unfortunately, at experimentally relevant
temperatures the finite size effects
on the cluster are greatly increased, so that
erratic behavior of the Drude weight as a function of
$\langle n \rangle$ and ${\rm U/t}$ is observed.
We estimate that for temperatures smaller than
$t/4 \approx 1200 \rm K$ our finite-cluster results are
not representative of the bulk limit.
It is thus more convenient to extrapolate
the experimental results to higher temperatures since the
slope $d\rho_{ab}/dT $ is accurately known experimentally.
These slopes (at the optimal doping concentration) are very
similar among the different cuprates, and range from
$d\rho_{ab}/dT \approx 1 \mu \Omega cm / K$
for ${\rm Bi_2 Sr_2 Cu O_6}$ (${\rm Bi2201}$) to
$d\rho_{ab}/dT \approx 0.5 \mu \Omega cm / K$
for ${\rm Y Ba_2 Cu_3 O_{7-\delta}}$ (Y123) with a T$_c$ of
90K. The extrapolated experimental results are shown in
Fig.2 (dotted lines). The theoretical predictions
obtained from the present cluster calculations are also
shown in this figure (open squares and triangles),
and were obtained by plotting $\rm D^{-1}$ times
a parameter with units of $\mu \Omega cm$, which sets
the relative scale between our calculations and experiment.
(Physically this parameter contains information about
scattering processes not incorporated in the Hubbard
model, so the overall normalization
of our predictions is not determined and has been taken
from experiment.)
Given this freedom to fix the overall normalization, good
agreement is observed between theoretical predictions and
extrapolated experimental results for $\rho_{ab}$ over the
range of temperatures for which we consider the cluster
results reliable.
This encouraging result suggests that the
simple one-band Hubbard model can describe several
normal state properties of the cuprates.
There is a slight upward curvature in the results,
which is not surprising since the experimentally measured
$\rho_{ab}$ is linear with temperature only at one particular
density.\cite{tak92} The filling fraction we have used,
$\langle n \rangle = 0.75$, may correspond to the slightly
``overdoped'' regime of the cuprates.\cite{other}

Now let us analyze the a.c. conductivity. To allow a
comparison with experiment we follow Imry\cite{imry} and
give each $\delta$-function of Eq.(1) a finite width,
to account for scattering through other processes not
included in the model, such as phonons and disorder. This
width $\epsilon$ should be larger than the mean inter-level
spacing in order to mimic a continuum of states, and thereby
produce dissipation.
$\epsilon^{-1}$ can be considered to be a phenomenological
relaxation time introduced to account for dissipative
processes not included in the Hamiltonian. $\epsilon$ is
a free parameter in our study (in addition to the electronic
density $\langle n \rangle$ and  the coupling ${\rm U/t}$ of
the Hubbard model) and we adopt $\epsilon = 0.33$ for the
following discussion.
In Fig.3,
$\sigma(\omega)$ is shown for the 8 site cluster
at several densities and couplings. The Drude peaks at
zero frequency are incorporated in the plots.
Fig.3a shows the result for a temperature comparable
to the antiferromagnetic exchange coupling J ${\rm (
T = 0.3125 t)}$, and for illustration we use
${\rm U/t = 20}$ to enlarge
the gap in the results. At half-filling (8 electrons),
most of the spectral weight is located at $\omega > 5$ (in t
units), in other words these are charge excitations,
as expected. As the density $\langle n \rangle$ is
decreased, spectral weight is transferred from the
high-frequency charge excitations to lower frequencies.
A Drude peak is formed, and considerable weight appears
within the insulating gap. (This may be associated with the
MIR band observed in the cuprates as has been discussed
extensively in the literature.\cite{review})
Fig.3b shows the same cluster at ${\rm T=1.25t}$ and
coupling ${\rm U/t = 8}$, which may be more
representative of the cuprates.\cite{review} Qualitatively,
the behavior is similar to that found at lower temperatures
and larger couplings, albeit with a smaller gap.
Little sub-structure is observed in the spectrum.
These results appear quite similar to the experimental
observations of Uchida et al. on ${\rm La_{2-x} Sr_x Cu O_4}$
(La214). Results for other high-T$_c$ cuprates are very similar.
Even the appearance of what Uchida et al\cite{uchida} called an
``isosbestic'' point (a point where conductivities for
different densities cross) is reproduced in this figure.
Finally, in the high temperature regime (${\rm T = 5t}$ in
Fig.3c), the gap is completely filled at all densities,
although a remnant of the upper Hubbard band can still be seen.
The MIR band and Drude peak have merged into a single structure.
In Fig. 3d, we show $\sigma(\omega)$ in the region $1< \omega < 5$
for $\rm U/t = 8$ and ${\rm T = 1.25 t}$. For the case of
$< n > = 0.5$, $\sigma(\omega)$ can be accurately described by a
$1/\omega^2$ law, as expected for a conventional Fermi liquid.
In contrast, for $< n > = 0.75$, $\sigma(\omega)$ has
a much slower decay with $\omega$ and can be fitted by the form
$(1/\omega^\alpha)$, with $\alpha = 1.3 \pm 0.2 $.
Both forms are included in this figure for comparison.
As can be seen in Fig.3b for 7 electrons ($< n > = 0.875$),
$\sigma(\omega)$ has an even slower decay with $\omega$ for
$\omega$ less than $\approx 3$.
This anomalous frequency dependency of the conductivity has also
been observed experimentally in La214.\cite{uchida}
Moreover, a close correlation between the temperature
dependence of $\rho_{ab}$ and the frequency dependence of the
scattering rate $1/\tau(\omega)$ (seen in the frequency
dependence of the conductivity) was observed;
$1/\tau(\omega) \sim \omega^{1.6}$ behavior in an overdoped
sample of La214 was recently reported,\cite{tak92} which is
quite reminiscent of our power law fit.

In summary, we have reported a numerical exact-diagonalization
calculation of the optical conductivity of the two dimensional
one-band Hubbard model. The d.c. resistivity
shows linear behavior in T for ${\rm T > t}$ and large
${\rm U/t}$, which presumably is also valid in the
lower-temperature range ${\rm J < T < t}$.
For these temperatures our results compare well with
experiment.
The a.c. conductivity was also calculated and we have presented
results at several temperatures. A MIR band is observed,
together with a Drude peak and charge excitations.
The combination of the Drude and MIR oscillator strengths
leads to a conductivity near half-filling that decays somewhat
more slowly than $1/\omega^2$ at
energies smaller than the insulator gap.

Upon completion of this work, we learned of an independent
study by Jakli\v{c} and Prelov\v{s}ek (University of Ljubljana
preprint, Dec. 1993) of the 2D t-J model at finite temperature
using a new numerical method. Their results are qualitatively
similar to our results for the Hubbard model.

Conversations with T. Barnes, M. B\"uttiker, E. Gagliano,
A. Moreo, and H. Pastawski
are gratefully acknowledged. J. R. was supported in part by
the U. S. Department of Energy (DOE) Office of Scientific
Computing, under the High Performance Computing and
Communications Program (HPCC), and in part by the DOE under
contract No. DE-AC05-84OR21400 managed by Martin Marietta
Energy Systems, Inc., and under
contract No. DE-FG05-87ER40376 with Vanderbilt University.
E. D. is supported by the Office of Naval Research under
grant ONR N00014-93-1-0495. The numerical calculations were done
at the Supercomputer Computations Research Institute,
Tallahassee, Florida and at the  National Center for
Supercomputing Applications, Urbana, Illinois.

\figure{ The inverse of the Drude weight, $\rm D^{-1}$, obtained
numerically on an 8-site cluster with APBC, as a function of
temperature (in units of the hopping parameter t).
The results are shown for couplings ranging from strong
(${\rm U/t = 20}$) to weak coupling (${\rm U/t = 4}$).
The filling fraction is shown in the figure.
\label{fig1}}

\figure{ The d.c. in-plane  resistivity $\rho_{ab}$ as a
function of temperature. The dotted lines correspond to
experimental results for Bi2201 and Y123, extrapolated to high
temperatures comparable to the hopping parameter (t is taken to
be 0.4 eV ($ \approx 4600 \rm K)$). The squares and triangles
are numerical results obtained on the 8-site cluster with APBC
for ${\rm U/t = 20}$ and filling fraction
$\langle n \rangle = 0.75$.
$\rm D^{-1}$ was multiplied by a constant with units of inverse
time to set the scale. The value of this constant was chosen
independently for the two compounds.
\label{fig2}}

\figure{ (a) The real part of the optical conductivity as a
function of frequency at different densities, for
${\rm U/t = 20}$, $\epsilon = 0.33$,
and ${\rm T = 0.3125t}$; (b)
as in (a) for ${\rm U/t = 8}$ and ${\rm T = 1.25t}$;
(c) as in (b) for ${\rm T = 5t}$;
(d) $\sigma(\omega)$ at ${\rm U/t = 8}$ and ${\rm T = 1.25t}$
in the interval $1 < \omega < 5$, together with fits to
$1/\omega^2$ ($<n> = 0.5$) and $1/\omega^{1.3}$
($<n> = 0.25$), indicated by dotted lines.
\label{fig3}}


\begin{references}
\bibitem{mc} Monte Carlo techniques require analytical
continuation from imaginary to real time to obtain
dynamical observables; therefore, this procedure should also
be compared to exact results obtained on finite clusters
to test its accuracy.
\end{references}
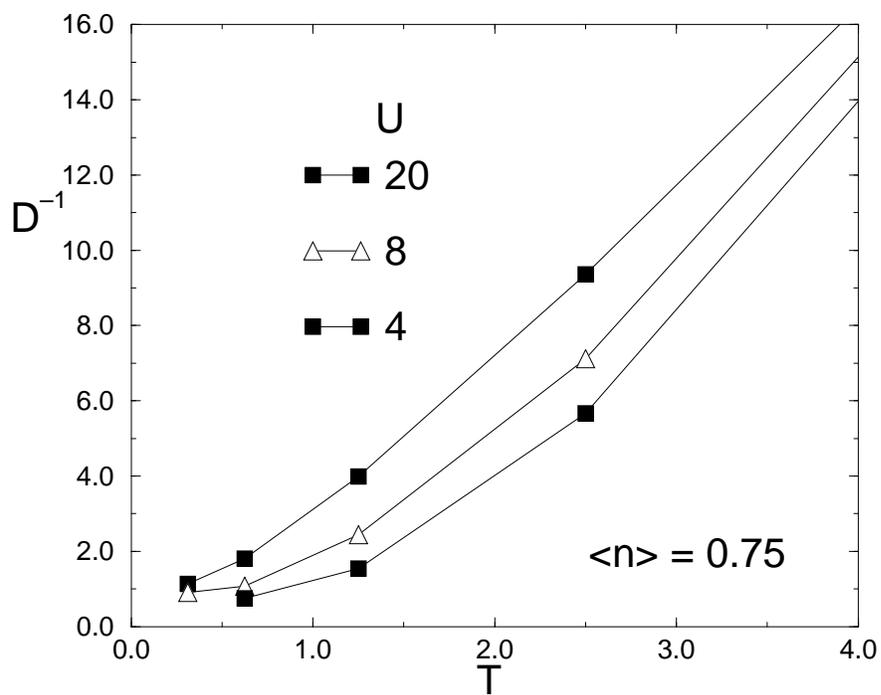
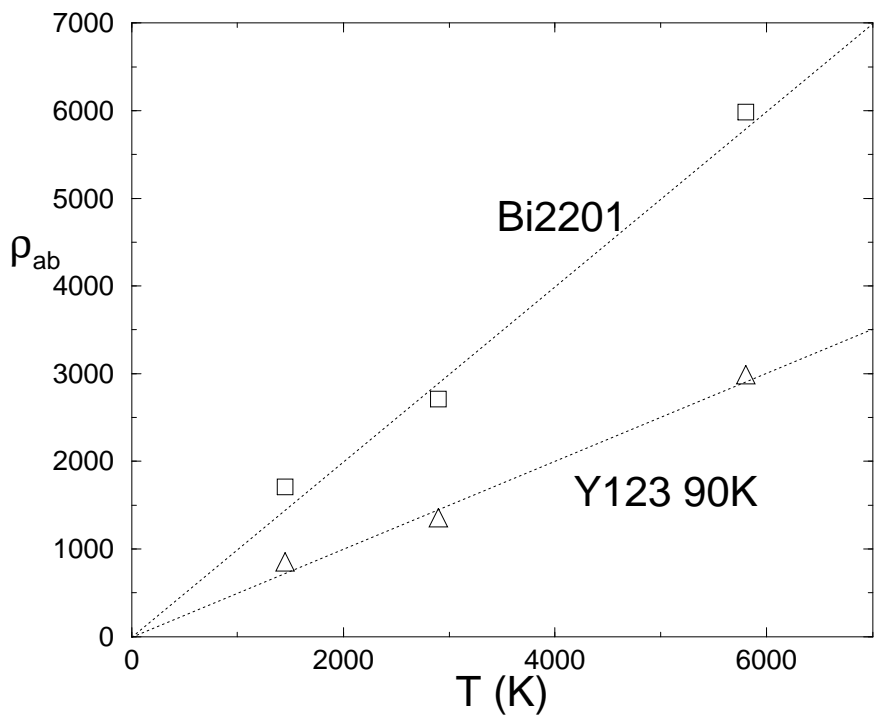
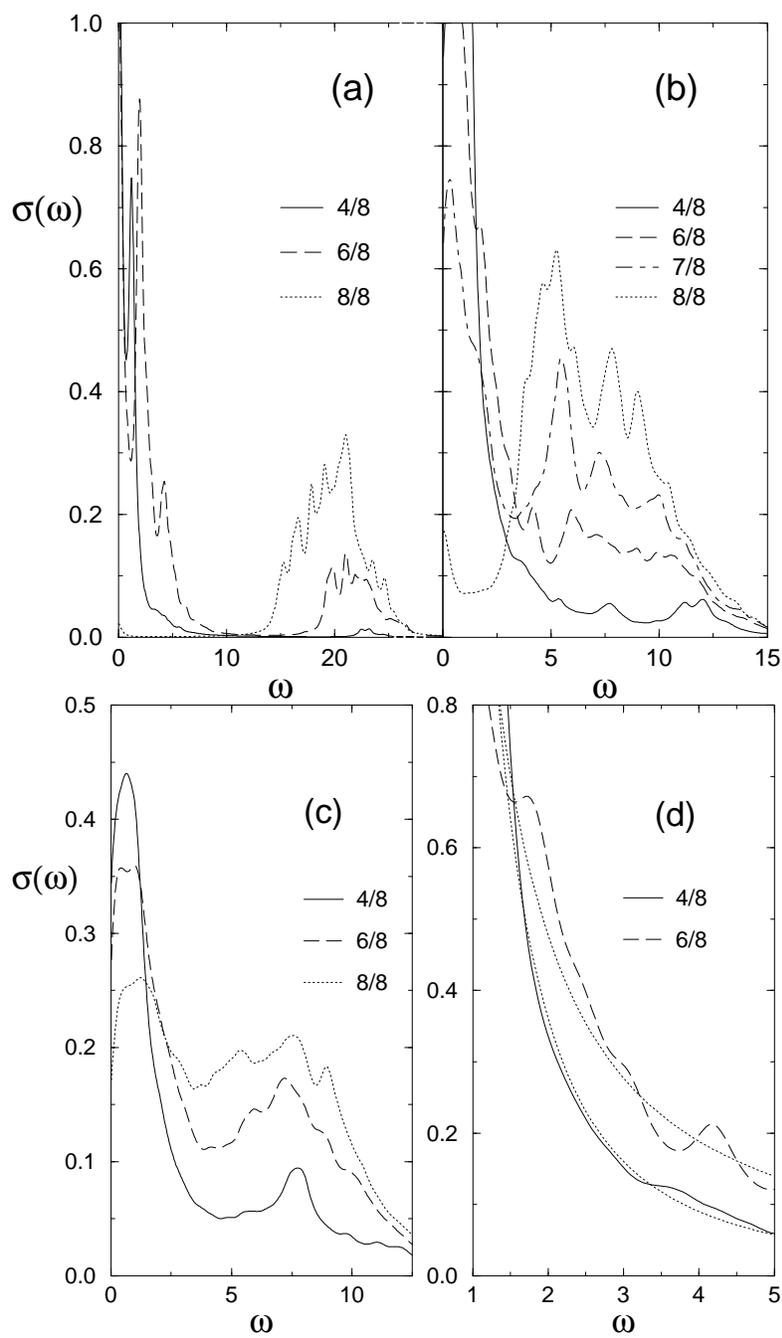
\end{document}